\documentclass[runningheads]{llncs}
\usepackage{graphicx}
\usepackage[misc]{ifsym}
\usepackage{multirow}
\usepackage{adjustbox}
\usepackage{xcolor}
\definecolor{faublue}{RGB}{0,56,101}
\begin{document}
\title{Automated Multi-sequence Cardiac MRI Segmentation Using Supervised Domain Adaptation}
\titlerunning{Automated Multi-sequence MRI Segmentation Using Domain Adaptation}
\author{Sulaiman Vesal\inst{1}(\Letter) \and Nishant Ravikumar\inst{1, 2} \and
Andreas Maier\inst{1}}

\authorrunning{S. Vesal et al.}
%First names are abbreviated in the running head.
%If there are more than two authors, 'et al.' is used.

\institute{Pattern Recognition Lab, Friedrich-Alexander-Universit\"at Erlangen-N\"urnberg, Germany\\ 
\and CISTIB, Centre for Computational Imaging and Simulation Technologies in Biomedicine, School of Computing, LICAMM Leeds Institute of Cardiovascular and
Metabolic Medicine, School of Medicine, University of Leeds, LS2 9JT,\\ 
United Kingdom\\
\email{sulaiman.vesal@fau.de}}

\maketitle              % typeset the header of the contribution
\begin{abstract}
Left ventricle segmentation and morphological assessment are essential for improving diagnosis and our understanding of cardiomyopathy, which in turn is imperative for reducing risk of myocardial infarctions in patients. Convolutional neural network (CNN) based methods for cardiac magnetic resonance (CMR) image segmentation rely on supervision with pixel-level annotations, and may not generalize well to images from a different domain. These methods are typically sensitive to variations in imaging protocols and data acquisition. Since annotating multi-sequence CMR images is tedious and subject to inter- and intra-observer variations, developing methods that can automatically adapt from one domain to the target domain is of great interest. In this paper, we propose an approach for domain adaptation in multi-sequence CMR segmentation task using transfer learning that combines multi-source image information. We first train an encoder-decoder CNN on T2-weighted and balanced-Steady State Free Precession (bSSFP) MR images with pixel-level annotation and fine-tune the same network with a limited number of Late Gadolinium Enhanced-MR (LGE-MR) subjects, to adapt the domain features. The domain-adapted network was trained with just four LGE-MR training samples and obtained an average Dice score of $\sim$85.0\% on the test set comprises of 40 LGE-MR subjects. The proposed method significantly outperformed a network without adaptation trained from scratch on the same set of LGE-MR training data. 

% To demonstrate the domain shift and robustness of our proposed method, we directly feed LGE-MR images to the first encoder-decoder which trained on T2+bSSFP only and it completely failed to segment ventricles and achieved an average Dice score of 33.3\%. 
% %  

\end{abstract}

\section{Introduction}
\begin{figure}[ht]
\centering
\includegraphics[width=10.0cm]{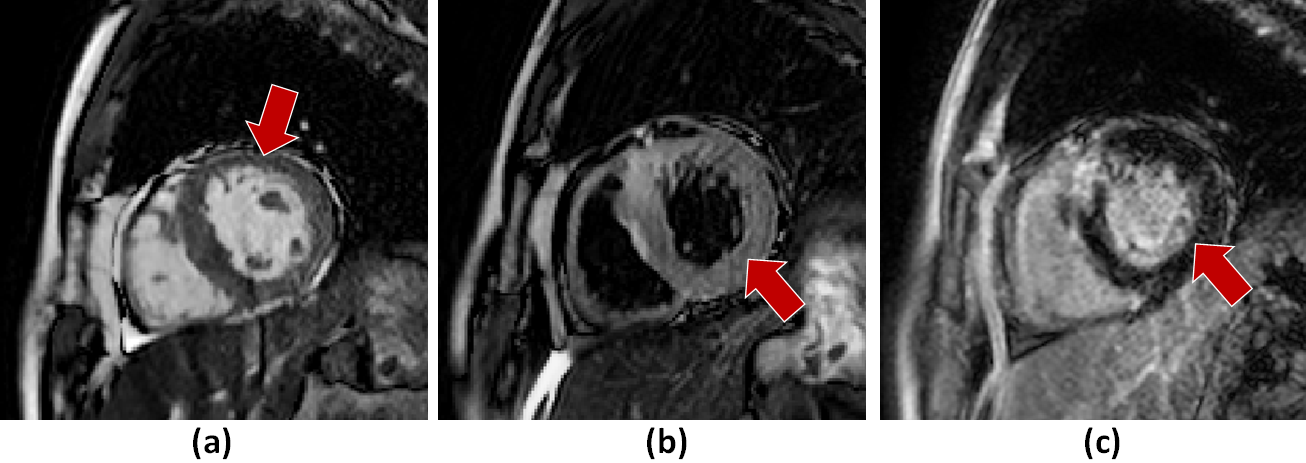}
\caption{Illustration of different CMR sequences: a) bSSFP slice, b) T2-weighted slice and c) LGE slice. The red arrows point to left ventricle on different sequences.} \label{fig2}
\end{figure}

Myocardial infarction (MI) is the leading cause of mortality and morbidity worldwide\cite{KIM20091}\cite{10.1093/eurheartj/ehx628}. Accurate analysis and modeling of the ventricles and myocardium from medical images are essential steps for diagnosis and treatment of patients with MI \cite{8458220}. MR imaging is used in the clinical workflow to provide anatomical and functional information of the heart. Different types of CMR sequences are acquired to provide complimentary information to each other, for example, T2-weighted images highlight the acute injury and ischemic regions, and the bSSFP cine sequence captures cardiac motion and presents clear boundaries. Moreover, LGE CMR can enhance the infarcted myocardium, appearing with distinctive brightness compared with healthy tissue \cite{10.1093/ehjci/jev123}. It is widely used to study the presence, location, and extent of MI in clinical studies. Thus, segmenting ventricles and myocardium from LGE CMR images is important to predict risk of infarcts, identify the extent of infarcted tissue and for patient prognosis \cite{kurzendorfer}. However, manual delineation is generally time-consuming, tedious and subject to inter- and intra-observer variations \cite{zhuang2}. In the medical image domain, heterogeneous domain shift is a severe problem, given the diversity in imaging modalities. For example, as shown in Fig. \ref{fig2}, cardiac regions visually appear significantly different in images acquired using different MR sequences. Generally, deep learning models trained on one set of MR sequence images perform poorly when tested on another type of MR sequence. One approach to maintain model performance in such a setting is to employ domain adaptation e.g. image to image translation or transfer learning. Domain adaptation attempts to reduce the shift between the distribution of data within the source and target domain.\\
\textbf{Related work.} Existing methods have approached multi-modal CMR segmentation using techniques such as cross-constrained shape \cite{LIU201949}, generative adversarial networks or 3D CNN. In \cite{10.1007/978-3-030-00919-9_17}\cite{Russo_2018_CVPR}, the authors first trained a CNN on the source domain and then transformed the target domain images into the appearance of source images, such that they could be analyzed using the network pre-trained on the source domain. However, these methods are based on generative adversarial networks and required substantial training data to achieve stable performance. On the other hand, there are limited works focusing on automatic LGE-CMR segmentation, which is a crucial prerequisite in a number of clinical applications of cardiology. Recently, few studies have attempted CMR multi-sequence segmentation. This type of method uses complementary information from multiple sequences to segment heart structures. \cite{zhuang2} proposed an unsupervised method using a multivariate mixture model (MvMM) for multi-sequence segmentation. MvMM adopted to model the joint intensity distribution of the
the multi-sequence images. The performance of this method depends on the quality of registration.\\
\textbf{Contributions.} In this study, we develop a deep learning-based method to segment the ventricles and myocardium in LGE CMR, combined with two other sequences (T2 and bSSFP) from the same patients. T2 and bSSFP sequences are used to assist the LGE CMR segmentation. Our method introduces a feature adaptation mechanism using transfer learning which explicitly adapts the features from T2 and bSSFP sequences to LGE images with few target training data. We first train an encoder-decoder CNN on T2 and bSSFP images with pixel-level annotation and re-train the same network with a limited number of LGE images, to adapt the learned features and imbue domain invariance between source and target domains. 

\section{Method}
\subsection{Domain Adaptation}
Deep learning methods are typically sensitive to domain shift and perform poorly on a new set of data with a different marginal probability distribution. However, annotating data for every new domain is a very expensive task, particularly in the medical area that requires clinical expertise. To segment LGE CMR images with very few annotated subjects, we attempted to get complementary information from other sequences with pixel-level annotations, and transfer the domain knowledge and initialize a second network with pre-trained weights. This is called as supervised domain adaptation \cite{8237871}.

Let's consider $D_{tb}$ as the image domain for T2+bSSFP sequences and $D_l$ for LGE sequences respectively. $D_{tb}$ can be expressed with feature space $S$ and associated probability distribution of $P(X)$ where $X = \{x_{1}, x_{2},...,x_{n}\} \in S$ \cite{5288526}\cite{10.1007/978-3-319-66179-7_59}. In a supervised learning task, domain $D_{tb} = \{S, P(X)\}$ consists of a model with associated objective function $\mathcal{F}_{tb}$, learning task of $T_{tb}$ and a label space of $Y$. The objective function $\mathcal{F}_{tb}$ for task segmentation $T_{tb}$ can be optimised using a pair of samples $\{x_i, y_i\}$ where $x_i \in X$ and $y_i \in Y$. After the training process, the learned model $\hat{\mathcal{F}_{tb}}$ can be used to predict on new samples of T2 and bSSFP images from $D_{tb}$ domain. Now, if we consider $D_l$ with LGE segmentation task $T_l$, we can transfer the learned weights from domain $D_{tb}$ to improve objective function of $F_l$ for segmenting LGE images in domain $D_l$ where $D_{tb}\neq D_l$ and $T_{tb} \neq T_l$. In this way, domain $D_{l}$ uses information from domain $D_{tb}$ to segment LGE images. The final model trained on T2+bSSFP domain and adapted to target domain (LGE) can be denoted as $\mathcal{F}_{lge}$.

\begin{figure}[ht]
\centering
\includegraphics[width=\textwidth]{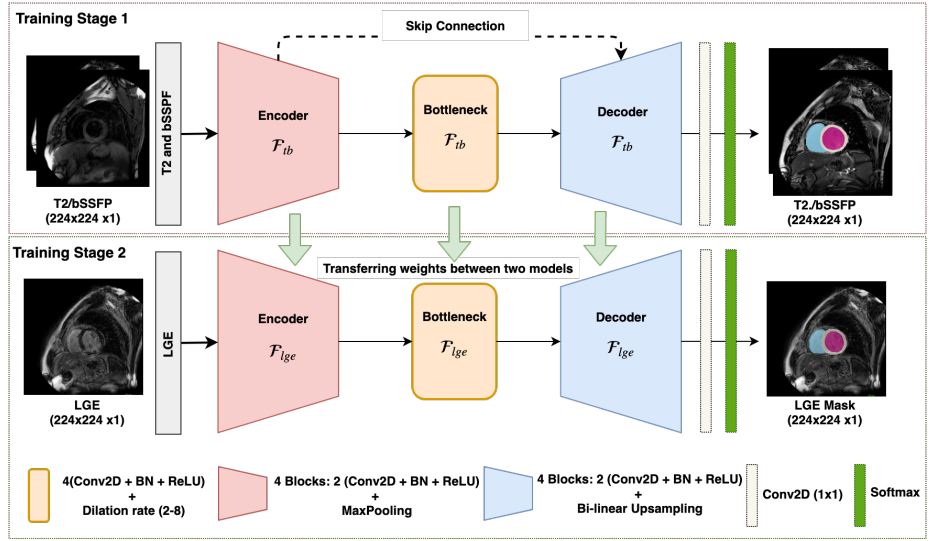}
\caption{Overview of our network architecture for feature transfer learning between different CMR sequences. The encoder-decoder first trained with T2+bSSFP images and in the second stage the network with learned weights retrained with LGE iamges.} \label{fig:3d_drunet}
\end{figure}

To construct the model $\mathcal{F}_{lge}$, we transferred the learned weights from $\mathcal{F}_{tb}$, then we retrain all the layers and fine-tuned the model on the limited training data from domain $D_l$. This is demonstrated in Fig. \ref{fig:3d_drunet}. All hyperparameters associated with the optimizer, the loss function, and the data augmentation scheme employed were kept the same for both models.

\subsection{Network Architecture}
Our network architecture is a fully convolutional network inspired from \cite{10.1007/978-3-030-12029-0_35} which comprises four encoder and decoder blocks, separated by a bottleneck block (refer to Fig. \ref{fig:3d_drunet}). The architecture includes skip connections between all encoder and decoder blocks at the same spatial resolution. Each encoder/decoder block consists of two 2D convolution layers, where, each convolution layer is followed by a batch-normalization and a Rectifier Linear Unit (ReLU) layer. In each encoder-convolution block, the input of the first convolution layer is concatenated with the output of the second convolution layer and zero-padded accordingly. The subsequent 2D max-pooling layer reduces the dimensions of the image by half. The use of residual connections \cite{resDeep} between convolution layers of each block in the encoder, help improve the flow of gradients in the backward pass of the network. The network utilizes a 1$\times$1 convolution to aggregate the feature maps from the final decoder block. This operation improves discriminative power as feature maps with lower activations are more likely to be suppressed through the assignment of lower weights. Finally, a softmax activation function was used in the last layer of the first network to classify the background from the foreground classes. Compared to U-Net, we replace the bottleneck convolution layers of the network with dilated convolutions \cite{Dilated} of size $3\times 3$, to enlarge the receptive field and enable the network to capture both local and global contextual information. The dilation rate of the four convolution layers is increased successively from $1-8$, and subsequently, their feature maps are summed together, enabling the network to capture the entire image's field of view. \\

\noindent\textbf{Multi-class Dice loss:}
To train the proposed network, a modified version of the soft-Dice loss is used which is less sensitive to class imbalance. This is motivated by the successful recent works \cite{10.1007/978-3-030-12029-0_35}\cite{7785132} for medical image segmentation. The Dice score is computed for each class individually, and then averaged over the number of classes. In order to segment an $N\times N$ input image (for example, a T1-weighted image with LV, RV, Myo and background as labels), the output of softmax layer is four probabilities for classes $k = 0, 1, 2, 3$ where, $\sum_{c}y_{n,k} = 1$ for each pixel. Given the one-hot encoded ground truth label $\hat{y}_{n,k}$ for that corresponding pixel, the multi-class soft Dice loss is defined as follows:
\begin{equation}
\label{eq15}
	\zeta_{dc}(y, \hat{y})  = 1- \frac{1}{K}(\sum_{k}\frac{\sum_{n}y_{nk} \hat{y}_{nk}}{\sum_{n}y_{nk} + \sum_{n}\hat{y}_{nk}})
\end{equation}

\subsection{Data Acquisitions}
We validated our proposed method on the STACOM MS-CMRSeg\footnote{http://www.sdspeople.fudan.edu.cn/zhuangxiahai/0/mscmrseg19/} 2019 challenge dataset with short-axis cardiac MR images of 45 patients diagnosed with cardiomyopathy. The dataset was collected in Shanghai Renji hospital with institutional ethics approval \cite{zhuang2}. Each patient had been scanned using three CMR sequences: LGE, T2, and bSSFP. Ground truth masks of cardiac structures were provided for 35 training samples (T2 and bSSFP only) and 5 validation samples, including the Left ventricle cavity (LV), the right ventricle cavity (RV), and the myocardium of the left ventricle (Myo). The LGE CMR was a T1-weighted, inversion-recovery, gradient-echo sequence, consisting of 10 to 18 slices covering the main body of the ventricles. The acquisition matrix was 512$\times$512, yielding an in-plane resolution of 0.75$\times$0.75 mm and slice thickness of 5 mm. The bSSFP CMR images consist of 8 to 12 contiguous slices, covering the full ventricles from the apex to the basal plane of the mitral valve, with some cases having several slices beyond the ventricles. These sequences have a slice thickness of 8-13 mm and an image resolution of 1.25$\times$1.25 mm. The T2-weighted CMR images consist of a small number of slices. Few cases have comprise just three slices, and the others have five (13 subjects), six (8 subjects) or seven (one subject) slices. The slice thickness is 12-20 mm and an in-plane resolution of 1.35$\times$1.35 mm. Since T1 and bSSFP images have very few slices, we combined both sequences together for the training of our backbone network.

\textbf{Preprocessing:} There is a large degree of variance in contrast and brightness across the MS-CMRSeg 2019 challenge images. The variability results from different system settings, and data acquisition which makes it harder for neural networks to process the images. Due to low contrast, we enhanced the image contrast slice-by-slice, using contrast limited adaptive histogram equalization (CLAHE). We normalized each MR volume individually to have zero mean and unit variance and cropped all images to 224$\times$224 to remove the black areas(background regions). Fig. \ref{fig2} shows MR slices of three different sequences of a patient after prepossessing. Furthermore, we use common training data augmentation strategies including random rotation, random scaling, random elastic deformations, random flips, and small shifts in intensity to increase training data. We employed the augmentation only on $x$ and $y$ axes and kept the volume depth the same. In this way, we do not degrade image quality.

\subsection{Network Training}
The organizers of the MS-CMRSeg challenge already split the data into training, validation and testing sets. In the first stage, we trained our model on 35 T2 and bSSFP sequences with pixel-level annotations and validated on 5 subjects using the adaptive moment estimation (ADAM) optimizer. For the second stage, we re-trained the model with five LGE subjects using 5-fold cross validation. The learning rate was fixed at $0.0001$, and the exponential decay rates of the $1$\textsuperscript{st} and $2$\textsuperscript{nd}-moment estimates were set to $0.9$ and $0.999$, respectively. During training, segmentation accuracy was evaluated on the validation set after each epoch of the network. Networks were trained until the validation accuracy stopped increasing, and the best performing model was selected for evaluation on the test set. The batch size for training T2+bSSFP images was set to 16 and during fine-tuning for LGE-MRI, to 4. We employed connected component (CC) analysis as a post-processing step to remove the small miss-classified regions in the output of the softmax layer. For inference, we use the weights which achieved the best Dice score on the validation set of LGE-CMR data set. The network was developed in Keras and TensorFlow, an open-source deep learning library for Python, and was trained on an NVIDIA Titan X-Pascal GPU with 3840 CUDA cores, and 12GB RAM.

\subsection{Evaluation Criteria}
To evaluate the accuracy of segmentation results, we used three different metrics to evaluate segmentation accuracy, namely, the Dice coefficient (Dice), Hausdorff distance (HD), and Average surface distance (ASD). The Dice metric measures the degree of overlap between the predicted and ground truth segmentation. It is the most widely used metric for evaluating segmentation quality in medical imaging. HD is defined as the maximum of the minimum voxel-wise distances between the ground truth and predicted object boundaries. ASD is the average of the minimum voxel-wise distances between the ground truth and predicted object boundaries. HD and ASD are evaluated using the shortest Euclidean distance of an arbitrary voxel $v$ to a point $P$, defined as $\bar{d}(v, P) = \min_{p\in P}||v-p||$.

% \begin{equation}
%     DSC (G, P) =  \frac{2 |G_{i} \cap P_{i}|}{|G_{i}| + |P_{i}|}
% \end{equation}
% where $G$ is the ground truth, and $P$ is the predicted mask respectively.  

% HD is defined as the maximum of the minimum voxel-wise distances between the ground truth and predicted object boundaries. This is expressed as:
% \begin{equation}
%     HD (G, P) = \max_{g\in G}\Big\{\min_{p\in P}\Big\{\sqrt{g^{2}-p^{2}}\Big\}\Big\}
% \end{equation}

% ASD is the average of the minimum voxel-wise distances between the ground truth and predicted object boundaries. By defining the shortest Euclidean distance of an arbitrary voxel $v$ to a point $P$ as $\bar{d}(v, P) = \min_{p\in P}||v-p||$, ASD can be expressed as:

% \begin{equation}
%     ASD (G, P) = \frac{1}{N_{G} + N_{P}}\Big\{\sum_{x_{p}\in P} \bar{d}(x_{P}, G) + \bar{d}(x_{G}, P) \Big\}
% \end{equation}
% where $N_P$ and $N_G$ are the number of voxels on the object boundaries in the predicted and ground truth masks, respectively.

\section{Results and Discussion}
\begin{figure}[ht]
\centering
\includegraphics[width=\textwidth]{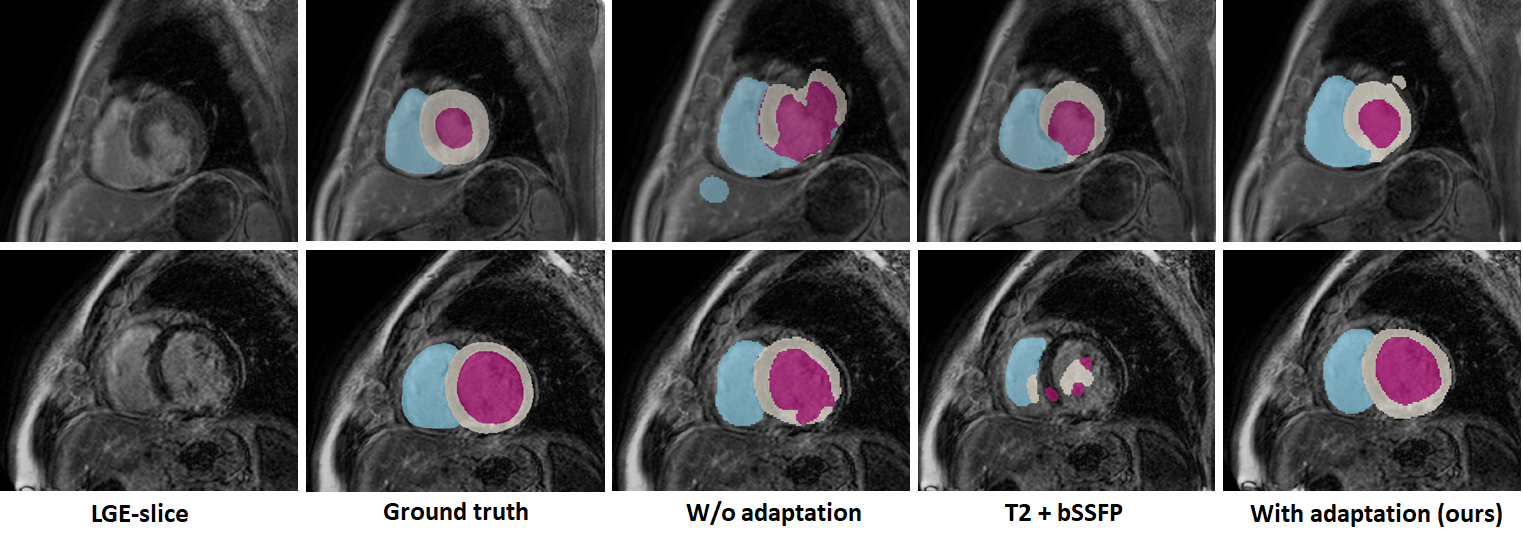}
\caption{Visual comparison of segmentation results produced by different methods. From left to right are the raw LGE-MR images, ground truth, W/o Adaptation output trained from scratch on LGE-MR images only, output from model trained on T2+bSSFP images, and proposed network. The cardiac structures of LV, RV, and Myo are indicated in purple, blue and withe color respectively. Each row corresponds to one subject.} \label{fig3}
\end{figure}

The proposed model is evaluated on the task of LGE-MRI segmentation. We compare our domain adapted network with three different training strategies including training without domain adaptation and predicting using the model trained on T2 and bSSFP only. Table \ref{tab:addlabel}. summarizes the comparison results, where we can see that our proposed method significantly improved the segmentation performance relative to networks tested without the adaptation strategy, in terms of Dice, HD, and ASD metrics. The model without domain adaptation and trained from scratch only on 5 LGE MR samples achieved an average Dice of 66.9\% on the validation set. Remarkably, for our proposed network with data augmentation, the average Dice improved to 80.9\% and HD and ASD was reduced to 13.6 and 1.0mm respectively. We achieved over 87.1\% Dice score for the LV structure and over 80.2\% Dice score for the RV. To illustrate the domain shift problem, we directly feed LGE-MR images to the first encoder-decoder after supervised training on T2+bSSFP domain. The result in Table 1 indicates that the source model completely failed on LGE-MR images with an average Dice score of 31.3\%, HD score of 37.37mm and ASD value of 11.06mm. Notably, compared with testing using model trained on T2+bSSFP, our method achieved superior performance especially for the LV and Myo structures, which are difficult to segment due to the presence of scars and blood pool within the cavity. Fig. \ref{fig3} illustrates segmentation results produced with different methods. Our method demonstrated robust segmentation performance on 40 LGE CMR sequences in the test dataset, summarized in Table \ref{tab:addlabel2}. We have achieved an average Dice score of 84.5\%  and HD, ASD score of 13.6 and 2.2mm respectively. The proposed model is able to reach a Dice score of 0.788 in terms of myocardium segmentation. These results help to highlight the generalization capacity of our approach to segmenting cardiac structures in LGE-MRI. Adapting the features between two domains results to a better model weight initialization and consequently improved the discrimination power of the second model. 

% Table generated by Excel2LaTeX from sheet 'Dice'
\begin{table}[ht]
\centering
\caption{Performance comparison between our proposed method and other segmentation methods for LGE-MR image segmentation on the validation data set.}
  \resizebox{\textwidth}{!}{\begin{tabular}{lcccc|cccc|cccc}
  \hline
  \multirow{2}[4]{*}{\textbf{Methods}} & \multicolumn{4}{c}{\textbf{Dice} $\uparrow$} & \multicolumn{4}{c}{\textbf{HD }[mm] $\downarrow$} & \multicolumn{4}{c}{\textbf{ASD }[mm] $\downarrow$} \\
\cline{2-13}        & Myo & LV & RV & Avg & Myo & LV & RV & Avg & Myo & LV & RV & Avg\\
\cline{1-13}      W/o adaptation & 0.527 & 0.775 & 0.705 & 0.669 & 57.23 & 129.67 & 124.13 & 103.68 & 2.96 & 1.37 & 1.35 & 1.89 \\
  T2+bSSFP & 0.169 & 0.386 & 0.383 & 0.313 & 42.62 & 30.37 & 39.13 & 37.37 & 14.6 & 7.04 & 11.53 & 11.06 \\
  W-adaptation & 0.671 & 0.862 & 0.766 & 0.766 & 17.6 & 13.25 & 22.78 & 17.88 & 2.77 & \textbf{0.75} & 3.81 & 2.44 \\
  W-adaptation+Aug & \textbf{0.749} & \textbf{0.871} & \textbf{0.802} & \textbf{0.807} & \textbf{11.35} & \textbf{15.66} & \textbf{14.05} & \textbf{13.69} & \textbf{1.06} & 0.81 & \textbf{1.18} & \textbf{1.02} \\
  \hline
  \hline
  \end{tabular}}%
\label{tab:addlabel}%
\end{table}%

\begin{table}[ht]
\centering
\caption{The Dice, Jaccard, HD and ASD score of domain-adapted method on the LGE CMR test dataset.}
  \begin{tabular}{p{2.4cm}p{2.3cm}p{2.3cm}p{2.3cm}p{2.3cm}}
  \hline
  \multirow{2}[4]{*}{\textbf{Structure}} & \multicolumn{4}{c}{Test-set results ($m \pm sd$)}\\
\cline{2-5}        & \textbf{Dice} $\uparrow$ & \textbf{Jaccard} $\uparrow$  & \textbf{ASD} [mm] $\downarrow$ & \textbf{HD} [mm]$\downarrow$ \\
  \hline
  Myo & 0.788 $\pm$ 0.073 & 0.656 $\pm$ 0.096 & 2.036 $\pm$ 0.616 & 12.53 $\pm$ 3.37 \\
  LV & 0.912 $\pm$ 0.033 & 0.840 $\pm$ 0.056 & 1.806 $\pm$ 0.615 & 11.29 $\pm$ 4.55 \\
  RV & 0.832 $\pm$ 0.084 & 0.721 $\pm$ 0.117 & 2.804 $\pm$ 1.376 & 17.11 $\pm$ 6.14 \\

  \hline
  \textbf{Average} & \textbf{0.844} $\pm$ \textbf{0.063} & \textbf{0.740} $\pm$ \textbf{0.090} & \textbf{2.215} $\pm$ \textbf{0.869} & \textbf{13.64} $\pm$ \textbf{4.68}\\
  \hline
  \hline
  \end{tabular}%
\label{tab:addlabel2}%
\end{table}%

\section{Conclusion}
In this study, we developed a robust deep learning approach for multi-sequence CMR image segmentation based on feature/domain-adaptation. Our network was first trained on T2-weighted and bSSFP sequences, and subsequently, the learned model weights were used to initialize the second network, and fine-tuned to segment LGE-MR images with a limited number of samples. The transfer leaning mechanism drastically reduces the domain shift during the training process. We validated our method on multi-sequence CMR images by comparing it with networks trained without domain adaptation. We employ our network on 2D slices as 3D models did not perform well on the MS-CMRSeg challenge 2019 dataset, given the limited number of axial slices. Experimental results highlight the advantage afforded by our approach, with regards to segmentation accuracy in LGE-MRI. Future work will aim to extend the framework using techniques for joint unsupervised image and feature adaptation using generative adversarial networks. \\

\noindent\textbf{Acknowledgments.} This work described in this paper was partially supported by the project EFI-BIG-THERA: Integrative ‘BigData Modeling’ for the development of novel therapeutic approaches for breast cancer. The authors would also like to thank NVIDIA for donating a Titan X-Pascal GPU.

\bibliographystyle{splncs04}
\bibliography{biblo.bib}
\end{document}